\documentstyle[floats,aps]{revtex}
\begin{document}
\input epsf
\draft

\twocolumn[\hsize\textwidth\columnwidth\hsize\csname
@twocolumnfalse\endcsname
\title{Superconducting Vortex with Antiferromagnetic Core}

\author{Daniel P. Arovas}
\address{Department of Physics, University of California at San Diego,
 La Jolla CA 92093}

\author{A. J. Berlinsky$^*$, C. Kallin$^*$, and Shou-Cheng Zhang}
\address{
Department of Physics, Stanford University, Stanford, CA 94305}

\date{\today}

\maketitle

\begin{abstract}
We show that a superconducting vortex in underdoped high $T_{\rm c}$
superconductors could have an antiferromagnetic core. This type of
vortex configuration arises as a topological solution in the recently
constructed SO(5) nonlinear $\sigma$ model and in Landau Ginzburg
theory with competing antiferromagnetic and superconducting 
order parameters. Experimental detection of this type of vortex 
by $\mu$SR and neutron scattering is proposed.
\end{abstract}

\vskip2pc]

\narrowtext

One of the most striking properties of high $T_{\rm c}$ superconductivity
is the close proximity between the antiferromagnetic (AF) and the 
superconducting (SC) phases. While there are a number of theories \cite{af}
linking the microscopic origin of high $T_{\rm c}$ SC to antiferromagnetic
correlations, it is natural to ask if the close proximity between these two
phases could have any macroscopic manifestations. Recently, a unified
theory\cite{so5} of AF and $d$-wave SC in the cuprates has been constructed
based on an SO(5) symmetry. In this theory, the AF and the $d$-wave SC order
parameters are unified into a five dimensional vector $(n_1,n_2,n_3,n_4,n_5)$
called a superspin. The AF order parameters $N_\alpha$ correspond to the
$(n_2,n_3,n_4)$ components, while the real and imaginary parts of the SC order
parameter $\Delta$ correspond to the $(n_1,n_5)$ components.  It was shown
that the chemical potential induces a first order superspin-flop transition
where the superspin abruptly changes direction from AF to SC.

The SO(5) theory predicts a spin triplet pseudo Goldstone boson associated
with the spontaneous breaking of SO(5) symmetry in the SC phase
\cite{demler,so5}; these can be identified with the recently observed resonant 
neutron scattering peaks in superconducting YBCO \cite{neutron}. Physically,
these modes corresponds to Gaussian fluctuations of the superspin.  However,
it was noted \cite{so5} that the SO(5) theory also admits a special class 
f topological solutions called meron configurations. In this configuration, 
the superspin lies inside the SC plane far away from the origin, and the SC
phase winds around the origin by $2\pi$. As the origin is approached 
from the radial direction, the superspin lifts up from the SC plane into
the AF sphere in order to minimize the energy cost of winding the SC phase.
The result is a SC vortex with an AF core \cite{vol}.

The existence of SC vortices with AF cores leads to non-trivial macroscopic
consequence which we shall explore in this paper. We first present detailed
analytic and numerical solutions of the SO(5) vortex. We also study a more
general Landau-Ginzburg (LG) theory obtained from the SO(5) theory by relaxing
the constraint on the magnitude of the superspin. This theory describes AF and
SC order parameters in competition with each other. We show that even if the
SC state wins in the bulk, under appropriate conditions a nonvanishing AF
component can occur inside a SC vortex core. The nature of the condition leads
us to conclude that a SC vortex with AF core should only be realized in
underdoped high $T_{\rm c}$ superconductors, not in the overdoped ones.  We
believe that the nature of the vortex core has nontrivial implication for the
physics  of high $T_{\rm c}$ superconductors in a high magnetic field.  In
recent experiments, Boebinger {\it et al.\/} \cite{boebinger} find that
insulating and normal phases appear upon destruction of SC by a high magnetic
field in underdoped and overdoped materials, respectively.
This observation could be intimately related to the
insulating/normal vortex core in the underdoped/overdoped materials which is
found in this work. We also suggest possible neutron and $\mu$SR experiments
to probe the AF components of the vortex core. 

The SO(5) theory has been constructed in its general form to allow for
anisotropy in the AF and SC couplings\cite{so5,burgess}.  However, in the
underdoped regime close to the AF-SC transition, most forms of anisotropies
are irrelevant\cite{burgess}. In this work, we first study the isotropic limit
of coupling constants, and allow only a quadratic symmetry breaking term.
In this limit of the SO(5) theory, the free energy density takes the form 
\cite{Zeeman}
\begin{eqnarray}
{\cal F}&=&{\textstyle{1\over 2}}\,\rho\,\bigg|\left({\vec\nabla}+
{i e^*\over\hbar c}{\vec A}\right)\psi\bigg|^2
-{\textstyle{1\over 2}}\,{\raise0.35ex\hbox{$\chi$}}\,(2\mu)^2\,|\psi|^2
+{\textstyle{1\over 2}}\,\rho\,|{\vec\nabla}{\vec m}|^2\nonumber\\
&&\qquad-{\textstyle{1\over 2}}\, g\,{\vec m}^2+{1\over 8\pi}
({\vec\nabla}\times{\vec A})^2\ ,
\label{Hden}
\end{eqnarray}
where $\psi=n^{\vphantom{\dagger}}_1+in^{\vphantom{\dagger}}_5$\ and
${\vec m}=n_2\,{\hat{\rm x}}+n_3\,{\hat{\rm y}}+n_4\,{\hat{\rm z}}$\ 
are the SC ($\psi$) and AF (${\vec m}$) order parameters.  Due to the
constraint $\psi^*\psi+{\vec m}^2=1$, the two are coupled.  When
${\tilde g}\equiv g-4\mu^2{\raise0.35ex\hbox{$\chi$}}\equiv
4{\raise0.35ex\hbox{$\chi$}}(\mu_{\rm c}^2-\mu^2)$ is negative, the bulk
phase is superconducting; ${\tilde g}>0$ prefers the antiferromagnet.
Assuming a constant direction for the N{\'e}el field,
we have ${\vec m}=\sqrt{1-|\psi|^2}\,{\hat m}$, and the equations
for $\psi$ are
\begin{eqnarray}
-\left({\vec\nabla}+{i e^*\over\hbar c}{\vec A}\right)^2\!\psi-\xi^{-2}\,\psi+
{{\vec\nabla}^2\sqrt{1-|\psi|^2}\over\sqrt{1-|\psi|^2}}\,\psi&=&0\nonumber\\
\lambda_{\scriptscriptstyle\rm L}^2\,{\vec\nabla}\times{\vec\nabla}\times
{\vec A}+|\psi|^2\,{\vec A}+ {\hbar c\over e^*}{1\over 2i}({\bar\psi}
{\vec\nabla}\psi - \psi{\vec\nabla}{\bar\psi})&=&0
\label{GLeqns}
\end{eqnarray}
where $\xi\equiv\sqrt{\rho/(-{\tilde g})}$ is the coherence length;
${\tilde g}<0$ in the superconducting phase.  As in the
more familiar Ginzburg-Landau theory of SC vortices, there
are two length scales in the problem.

\begin{figure} [t]
\centering
\leavevmode
\epsfxsize=8cm
\epsfysize=8cm
\epsfbox[18 144 592 718] {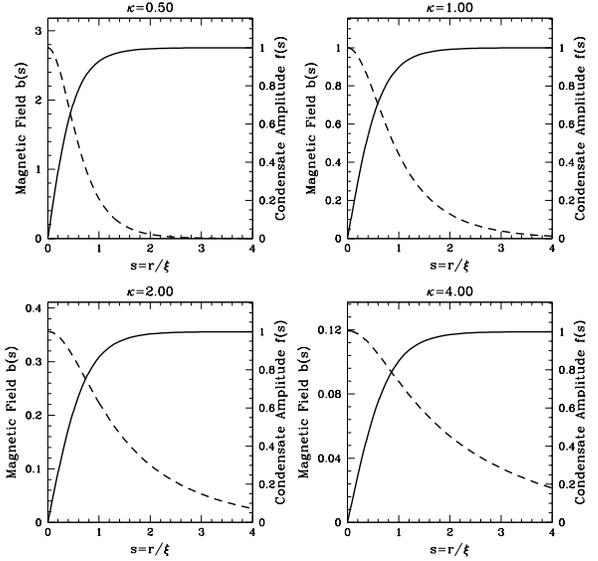}
\caption[]
{\label{vortex} Vortex profiles for a series of different values of $\kappa$.
The dashed lines show the magnetic field distribution and the solid
lines show the SC order parameter.  The AF order parameter is related
to the SC order parameter by the SO(5) constraint.}
\end{figure}

{\it Vortex Solutions} -- In searching for vortex solutions, it is convenient
to work in polar coordinates $(r,\phi)$ and to rescale distance,
$r\equiv\xi\,s$,and vector potential, ${\vec A}({\vec r})\equiv
(\phi_{\scriptscriptstyle\rm L}/2\pi\xi^2)\,s^{-1}\alpha(2)\,{\hat\phi}$, where
${\phi_{\scriptscriptstyle\rm L}}=hc/e^*$ is the London flux quantum.  The
magnetic field is then $B(r)=({\phi_{\scriptscriptstyle\rm L}}/2\pi\xi^2)\,b(s)$
with $b(s)=s^{-1}\,d\alpha/ds$.  We demand $\alpha(0)=0$ and
$\alpha(\infty)=m$, the number of flux quanta through the plane.  With
$\psi=f(r)\exp(im\phi)$, then,
\begin{eqnarray}
0&=&{d^2\!f\over ds^2}+{1\over s}{df\over ds}
+{f\over{ 1 - f^2}} \left({df\over ds}\right)^2\nonumber\\
&&\quad+\left[1-\left(\alpha- m\over s\right)^2\right](1-f^2)f\nonumber\\
0&=&{d^2\! \alpha\over ds^2}-{1\over s}{d\alpha\over ds}
-{(\alpha -m)\over\kappa^2}\,f^2
\label{GLprime}
\end{eqnarray}
where $\kappa\equiv\lambda_{\scriptscriptstyle\rm L}/\xi$ is the
Ginzburg-Landau parameter.  We solve these equations by the shooting method,
using the asymptotic solutions $f(s\to 0)\sim C_1\,s^m$,
$\alpha(s\to 0)\sim C_2\,s^2$, $f(s\to\infty)\sim 1-C_3\,\exp(-2s)$,
$\alpha(s\to\infty)\sim m-C_4\,\exp(-2\sqrt{s}/\kappa)$,
where $C_1,\ldots,C_4$ are constants.  The SC order parameter and magnetic
field distribution near the vortex core is shown in Fig. \ref{vortex} for
different values of $\kappa$.  In the SO(5) theory, the AF order parameter
profile is simply given by $\sqrt{1-\Psi^2}$.  As in conventional vortex
solutions, the SC order parameter is well approximated by a $\tanh$.
The vortex line energy has also been calculated and gives a value of
$\kappa_{\rm crit}$, separating type I and type II behavior, in good agreement 
with the value predicted from $H_{{\rm c}2}$ for this model,
$\kappa_{\rm crit}=1$ \cite{unpub}.

\begin{figure} [t]
\centering
\leavevmode
\epsfxsize=8cm
\epsfysize=8cm
\epsfbox[18 144 592 718] {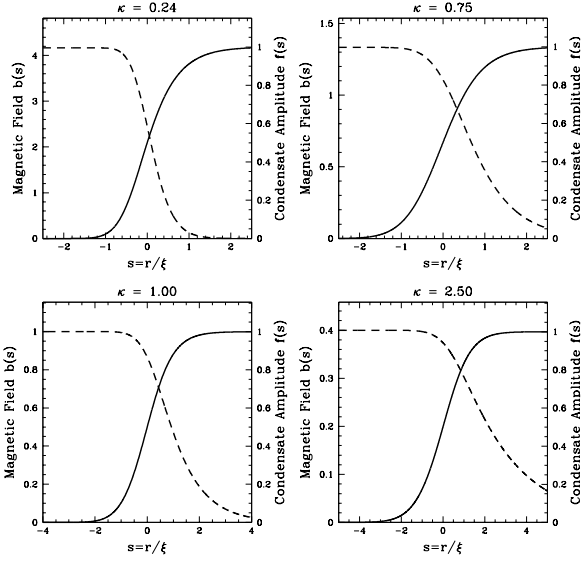}
\caption[]
{\label{dwall} Domain wall profiles for a series of different values
of $\kappa$.  The dashed lines show the magnetic field distribution
and the solid lines show the SC order parameter.}
\end{figure}

{\it Domain Walls} -- The Gibbs free energy density is ${\cal G}={\cal F}-
{1\over 4\pi}{\vec B}\cdot{\vec H}$.  In the bulk, the SC state is
characterized by $|\psi|=1$, ${\vec B}=0$, and a free energy density of
${\cal G}_{\scriptscriptstyle\rm SC}={\textstyle{1\over 2}}{\tilde g}=
-\rho/2\xi^2$.  The AF state has $|\psi|=0$, ${\vec B}={\vec H}$, and
${\cal G}_{AF}=-{\vec H}^2/8\pi$.  Setting ${\cal G}_{\scriptscriptstyle\rm SC}
={\cal G}_{AF}$ gives the thermodynamic critical field $H_{\rm c}$:
$H_{\rm c}=\sqrt{4\pi\rho}/\xi=\phi_{\scriptscriptstyle\rm L}/2\pi\xi
\lambda_{\scriptscriptstyle\rm L}$.  We now consider a domain wall separating
a bulk AF $(x\to-\infty)$ from a bulk SC $(x\to+\infty)$ and compute the energy
of the domain wall relative to that of either bulk state.  We write $x=\xi\,s$
and ${\vec A}=(\hbar c/e^*\xi)\,a(s)\,{\hat {\rm y}}$ to obtain the
Ginzburg-Landau equations
\begin{eqnarray}
0&=&{\partial^2\!f\over\partial s^2}+(1-a^2)\,f\,(1-f^2)+{f\over 1-f^2}
\left({\partial f\over\partial s}\right)^2\nonumber\\
0&=&{\partial^2\!a\over\partial s^2}-{1\over\kappa^2}\,a\,f^2
\label{DWEL}
\end{eqnarray}
with asymptotic solutions $f(s\to -\infty)$ $\sim$ $C_1\,\exp(-s^2/2\kappa)$,
$a(s\to -\infty)$ $\sim$ $s/\kappa+C_2$, $f(s\to\infty)$ $\sim$ $1-C_3\,
\exp(-2s)$, and $a(s\to\infty)$ $\sim$ $C_4\,\exp(-s/\kappa)$.  We again solve
by the shooting method and the results are shown in Fig. \ref{dwall} for
various values of $\kappa$.

The domain wall free energy per unit length is
\begin{equation}
\sigma={\rho\over 2\xi}\int_{-\infty}^\infty\!\!\!ds\left\{
\left(\kappa\,{\partial a\over\partial s}-1\right)^2-\left({f\over 1-f^2}\,
{\partial f\over\partial s}\right)^2\right\}\ .
\label{DW2}
\end{equation}
Note that the integrand of equation (\ref{DW2}) is a difference of two positive
quantities.  When $\sigma> 0$ the domain wall energy is  positive. 
This is type I behavior.  When $\sigma <0$ the domain wall energy is negative  
and we have type II behavior.  $\sigma =0$ corresponds to
$\kappa_{\rm crit}=0.30$, which differs from that determined by $H_{{\rm c}2}$
or the vortex line energy because of the gradient which appear in the fourth
order terms due to the SO(5) constraint.\cite{unpub}

In the above calculations, the SO(5) constraint forces the vortex core
to be antiferromagnetic.  A normal core is describable within a soft
superspin model.  To explore the competition between AF and normal cores,
we write $\psi=n\cos\theta \,e^{i\phi}$ and ${\vec m}=n\sin\theta\,{\hat m}$
with $\phi=\tan^{-1}(y/x)$.  We further assume that ${\hat m}$ is constant
and we work in the extreme type-II limit where the magnetic field is ignored.
The free energy density is then
\begin{eqnarray}
{\cal F}&={\textstyle{1\over 2}}\rho\left[({\vec\nabla}n)^2+
n^2({\vec\nabla}\theta)^2+{1\over r^2}\,n^2\cos^2\theta\right]\nonumber\\
&\qquad+{\textstyle{1\over 2}} a n^2 +{\textstyle{1\over 4}}{\tilde g}
(\cos 2\theta -1)n^2+{\textstyle{1\over 4}}b n^4
\label{soft}
\end{eqnarray}
where $a(T)=a' (T-T_{\rm c})$ in the vicinity of $T_{\rm c}$.
We now consider two trial vortex profiles: ({\it i\/}) $n(r)=n_0\tanh(r/\xi)$,
$\theta(r)=0$ (normal core), and ({\it ii\/}) $n(r)=n_0$,
$\theta(r)={\textstyle{\pi\over 2}}\exp(-r/\ell)$ (antiferromagnetic core).
Here, $\xi$ and $\ell$ are variational parameters, while $n_0=\sqrt{(-a)/b}$
is the superspin magnitude far from the vortex core.  We find
$\xi^2=3.385\,\rho/(-a)$ and $\ell^2=0.9865\,\rho/(-{\tilde g})$.
The difference in free energy per unit length is then
$F_{\scriptscriptstyle\rm AF}-F_{\rm normal}\equiv\pi\rho\, n_0^2\,
(X(\lambda)-X_{\rm c})$, where $\lambda=1.852\sqrt{{\tilde g}/a}$,
$X_{\rm c}=0.3214$, and
\begin{equation}
X(\lambda)=\int_0^\infty {du\over u}\,\left[\tanh^2 u
-\cos^2({\textstyle{\pi\over 2}} e^{-\lambda u})\right]\ .
\end{equation}
We find that the AF core is preferred for $\lambda < 0.5683$, {\it i.e.\/}
${\tilde g}/a < 0.0941$.

The consequence of this analysis is that there is a line in the $(T,\mu)$
plane separating regions with antiferromagnetic and normal cores.  We find
AF cores stable for $4{\raise0.35ex\hbox{$\chi$}}(\mu^2-\mu_{\rm c}^2)
< 0.0941 (T_{\rm c}-T)$, {\it hence, AF cores should be observable in
underdoped materials at low temperatures}.  Increasing doping or
temperature will eventually result in normal cores.

The general form of the free energy density when SO(5) has been broken down to
O(2)$_{\scriptscriptstyle \rm SC}\times$O(3)$_{\scriptscriptstyle \rm AF}$ is
\begin{eqnarray}
{\cal F}&={\textstyle{1\over 2}}\rho^{\vphantom\dagger}_\psi
|{\vec\nabla}\psi|^2+
{\textstyle{1\over 2}}\rho^{\vphantom\dagger}_m|{\vec\nabla}{\vec m}|^2 
+{\textstyle{1\over 2}}\alpha\,|{\vec m}|^2 +
{\textstyle{1\over 2}}\beta\,|\psi|^2 \nonumber\\
&\qquad + {\textstyle{1\over 2}} u\,|{\vec m}|^4+
w\,|{\vec m}|^2\,|\psi|^2+{\textstyle{1\over 2}}v\,|\psi|^4\ ,
\label{LG1}
\end{eqnarray}
where we again work in the extreme type-II limit.  We take $\beta < 0$ and
$u,v,w >0$.  Bulk SC is stable if $\alpha/\beta < w/v$; this also
precludes a mixed ($|\psi|, m\neq 0$) phase.  If $\alpha < 0$ and
$\alpha/\beta > u/w$ we must impose $\beta^2/v > \alpha^2/u$ for global SC
stability. We write $\psi({\vec r})=\sqrt{|\beta|/2v}\,f(s)\,e^{i\phi}$ with
$s\equiv ({|\beta|/\rho^{\vphantom\dagger}_\psi})^{1/2}\,r$;
$f$ is given by the solution to
\begin{equation}
{d^2\!f\over ds^2}+{1\over s}{df\over ds}+\left(1-{1\over s^2}\right)f-f^3=0
\end{equation}
subject to $f(0)=0$ and $f(\infty)=1$.
The linearized equation for $m(s)$ is then
\begin{equation}
-{\rho^{\vphantom\dagger}_m\over \rho_\psi}
\left({d^2\!m\over ds^2}+{1\over s}
{dm\over ds}\right)+{w\over v}\,f^2(s)\,m={\alpha\over\beta}\,m\ .
\end{equation}
This defines an eigenvalue problem, perhaps conveniently considered as
a radially symmetric Schr{\"o}dinger equation for a particle of mass
$M=\hbar^2\rho^{\vphantom\dagger}_\psi/2\rho^{\vphantom\dagger}_m$
in an attractive potential $V(s)=-(w/v)(1-f^2(s))$; the energy eigenvalue is
$E=(\alpha/\beta-w/v)$.  Bound states, for which $m(\infty) = 0$, satisfy
$E<0$, in agreement with the aforementioned conditions.  Antiferromagnetic
cores will exist for $E>-\Upsilon$, where $-\Upsilon$ is the lowest bound
state energy; clearly $\Upsilon(\rho^{\vphantom\dagger}_\psi/
\rho^{\vphantom\dagger}_m,w/v)$ is an increasing function of $w/v$ which
vanishes when $w/v=0$. Thus, we arrive at the condition
$w/v-\Upsilon(\rho^{\vphantom\dagger}_\psi/\rho^{\vphantom\dagger}_m,w/v)
< \alpha/\beta < w/v$.  To compare with our earlier variational calculation,
set $\alpha/\beta=1-{\tilde g}/a$ and $\rho^{\vphantom\dagger}_\psi/
\rho^{\vphantom\dagger}_m=w/v=1$.

\begin{figure} [!t]
\centering
\leavevmode
\epsfxsize=8cm
\epsfysize=8cm
\epsfbox[18 144 592 718]{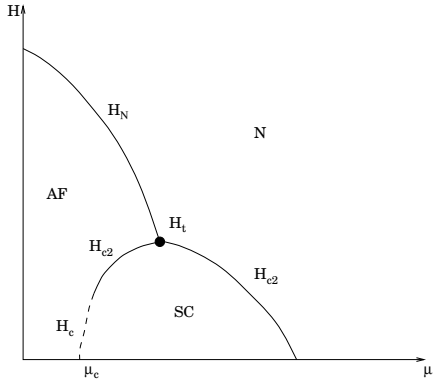}
\caption[]
{\label{phase} Schematic $T=0$ phase diagram}
\end{figure}

The SC vortex with an AF core has important consequences for the
high magnetic field physics in underdoped high $T_{\rm c}$ 
superconductors.  Within the SO(5) theory, both the thermodynamic critical
field  $H_{\rm c}={\phi_{\scriptscriptstyle\rm L}}/ 2\pi\xi
\lambda_{\scriptscriptstyle\rm L}$ and the upper critical field
$H_{{\rm c}2}={\phi_{\scriptscriptstyle\rm L}}/2\pi\xi^2$
describe phase transitions between SC and AF phases {\it at fixed
chemical potential $\mu$}.  A schematic zero temperature phase diagram in
the $(H,\mu)$ plane is shown is Fig. \ref{phase}.  Applying a uniform magnetic
field to the AF causes the Neel vector to flop into the plane perpendicular to
the applied field, while the total magnetization vector is aligned in the
field direction.  The bulk AF-to-normal transition occurs at a critical
field $H_{\scriptscriptstyle\rm N}=\alpha J/\hbar\mu_B$ (about 50 Tesla
if $\alpha=1$), where $J$ is the AF exchange constant and $\alpha$ is a
dimensionless constant.  Since doping (increasing $\mu$) significantly
weakens $J$, we expect $H_{\scriptscriptstyle\rm N}$ to decrease with
increasing $\mu$.  On the overdoped SC side, we also expect
$H_{{\rm c}2}$ to decrease with increasing doping because of the loss in
pairing energy.  Surprisingly, on the underdoped side, the SO(5) theory gives
$H_{{\rm c}2}=4{\raise0.35ex\hbox{$\chi$}}{\phi_{\scriptscriptstyle\rm L}}
(\mu^2-\mu_{\rm c}^2)/2\pi\rho$ which increases with $\mu$.  These
three critical lines meet at a common tricritical point $H_{\rm t}$.
Several important features are to be noted about our proposed phase diagram.
First of all, if we assume that the London penetration depth
$\lambda_{\scriptscriptstyle\rm L}$ remains finite at $\mu_{\rm c}$, then
$H_{\rm c}$ will exceed $H_{{\rm c}2}$ sufficiently close to $\mu_{\rm c}$,
since $\xi^{-2}$ behaves as $\mu^2-\mu_{\rm c}^2$.  Therefore,
the SC-to-AF transition will change from second order to first order
in the vicinity of $\mu_{\rm c}$, where these two phases are separated by the
thermodynamic field $H_{\rm c}\propto \sqrt{\mu -\mu_{\rm c}}$.  Secondly,
all our discussions are carried out for a short ranged model at fixed $\mu$;
Coulomb interactions may lead to a significant modification of
the phase diagram.  However, we believe the most salient feature of our
phase diagram, namely the transition from a SC state to an insulating
state with applied field, will still remain valid in the underdoped regime.
This could offer a basic explanation of the observation
by Boebinger {\it et al.\/} \cite{boebinger}  

Next we consider possible experimental methods for observing AF vortex cores.
The AF vortex core size is on the order of the SC coherence length (four to
five lattice constants).  The density of vortices is proportional to the field.
The local electron magnetic dipolar fields in the cores are hundreds of Gauss,
as they are in the pure AF \cite{muon}, but because
these AF regions are at best one-dimensional (along the field direction),
the local fields may not be static, except perhaps at low temperatures. 
The correlations along these small AF tubes depend on the
weak interlayer exchange.  Spin correlations between AF cores of
neighboring vortices will be even weaker, and hence 
the AF order may not be coherent from one vortex to the next.
This will show up in the neutron diffraction pattern.
For neutron scattering, one expects to see peaks around $(\pi,\pi)$ with a
width of (core size)$^{-1}$.  If the spins are not static, this scattering
will be quasi-elastic.  The total intensity can be obtained by integrating
over low frequencies.  If the AF order were coherent from one vortex to
the next, the periodicity of the vortex lattice would lead to superlattice
diffraction peaks split by (inter-vortex spacing)$^{-1}$.  In any case, the 
integrated intensity of the broad peak at $(\pi,\pi)$ will scale with field.
For transverse field $\mu$SR, if the AF fields are static on the time
scale of the muon precession frequency, it may be possible to observe the
staggered local electron dipole field directly.  The reason is that the
field distribution from the center of a normal vortex core appears as a
step at the high field end of the $\mu$SR spectrum.  For AF cores, the
dipolar field at the muon site will have a longitudinal component (along
the $c$-axis), of order 100 G \cite{muon}, even if the electron spins lie in
the $a$-$b$ plane.  This longitudinal field at the muon site or sites will have 
random sign and hence generate double or multiple steps in the absence of
other broadening mechanisms.  In a longitudinal field experiment, the
fluctuating transverse fields due to the AF vortex cores could give rise to a
substantial amount of relaxation.  The magnitude of the relaxation depends on
the amount of spectral density of transverse spin fluctuations at the Larmor
frequency of the muon.  The temperature dependence of this relaxation is a
probe of the spin fluctuations within the vortex cores.

{\it Acknowledgments} -- The authors are indebted to E. Demler for
numerous valuable discussions and to N. Andersen, J. Brewer, P. Hedeg\aa rd,
B. Keimer, R. Kiefl and T. Mason for very useful discussions and correspondence
regarding possible experiments.  DPA is grateful to H. Levine for
discussions on shooting.  The financial support of NSF
grants DMR-9400372, DMR-9522915 and of NSERC is gratefully acknowledged.


\begin{thebibliography}{1}
\bibitem{Zeeman} We ignore the Zeeman coupling, given by
${\cal H}_{\rm Zeeman}=-{\textstyle{1\over 2}}\,{\raise0.35ex\hbox{$\chi$}}\,
(\gamma B)^2\,(m_1^2+m_2^2)$.
\end{thebibliography}
\end{document}